\documentclass[12pt,preprint]{aastex} 
\usepackage{emulateapj5}

\shorttitle{Preheating and the $M_{gas} - T_X$ Relation}
\shortauthors{McCarthy, Babul, \& Balogh} 
\submitted{To appear in the Astrophysical Journal (received 01/22/02, 
accepted 03/13/02)}
\journalinfo{astro-ph/0203189}

\begin{document} 

\title{\bf The Cluster $M_{gas} - T_X$ Relation: Evidence for 
a High Level of Preheating}

\author{Ian G. McCarthy$^{1,2}$, Arif Babul$^1$, \& Michael L. Balogh$^3$}

\affil{$^1$Department of Physics \& Astronomy, University of Victoria, Victoria, BC, V8P 
1A1, Canada}

\affil{$^3$Department of Physics, University of Durham, South Road, Durham, DH1 3LE, UK}

\footnotetext[2]{email: mccarthy@beluga.phys.uvic.ca}

\begin{abstract} 

Recent X-ray observations have been used to demonstrate that the cluster gas mass - 
temperature relation is steeper than theoretical self-similar predictions drawn from 
numerical simulations that consider the evolution of the cluster gas through the 
effects of gravity and shock heating alone.  One possible explanation for this is that 
the gas mass fraction is not constant across clusters of different temperature, as 
is usually assumed.  Observationally, however, there is no compelling evidence for gas 
mass fraction variation, especially in the case of hot clusters.  Seeking an 
alternative physical explanation for the observed trends, we investigate the role in 
the cluster gas mass - temperature relation of the preheating of the intracluster 
medium by some arbitrary source for clusters with emission-weighted mean temperatures 
of $T_X \gtrsim 3$ keV.  Making use of the physically-motivated, analytic models 
developed in 2002 by Babul and coworkers, we find that preheating does, indeed, 
lead to a steeper relation.  This is in agreement with previous theoretical studies 
on the relation.  However, in apparent conflict with these studies, we argue that a 
``high'' level of entropy injection is required to match observations.  In 
particular, an entropy floor of $\gtrsim 300$ keV cm$^2$ is required.  We also 
present a new test, namely, the study of the relation within different fixed radii.  
This allows one to indirectly probe the density profiles of clusters, since it samples 
different fractions of the virial radius for clusters of different temperature.  This 
test also confirms that a high level of preheating is required to match observations.

\end{abstract}


\section{Introduction}

Analytic models and numerical simulations of clusters of galaxies have been used to 
predict the existence of scaling relations between various observable quantities, such 
as the well-known luminosity ($L_X$) -  temperature ($T_X$)  and mass ($M$) -  
temperature relations, where \hbox{$L_X$ $\propto$ $T_X^2$} and \hbox{$M$ $\propto$ 
$T_X^{1.5}$}, respectively.  However, it is now fairly well established that X-ray 
properties of clusters do not scale in such a fashion.  Most notable of these is the 
\hbox{$L_X - T_X$} relationship, which is observed to be much steeper than predicted, 
\hbox{$L_X$ $\propto$ $T_X^{2.6-3.0}$} (e.g., Markevitch 1998; Allen \& Fabian 
1998; Arnaud \& Evrard 1999).

Considerable effort has recently been directed towards explaining why the observed 
relations deviate from their predicted scalings (e.g., Tozzi \& Norman 2001; Dav\'{e} 
et al. 2001; Babul et al. 2002, hereafter BBLP02).  In particular, it is the 
$L_X - T_X$ relation that has grabbed most of the spotlight because there is a wealth 
of published observational studies on the luminosities and temperatures of clusters 
with which to compare models and simulations.  However, another important scaling 
relation is the cluster gas mass \hbox{($M_{gas}$) - $T_X$} relation.  Neumann \& 
Arnaud (2001) have suggested that a deviation from the self-similar scaling of 
$M_{gas} \propto T_X^{1.5}$ might ``explain'' the observed deviation in the $L_X - 
T_X$ relation.  Indeed, a number of observational studies have indicated that the 
relation is much steeper, with $M_{gas} \propto T_X^{1.6-2.0}$ (Vikhlinin et al. 
1999; Mohr  et al. 1999, hereafter MME99; Neumann \& Arnaud 2001).  If the gas 
density profile is roughly self-similar, this does lead to consistency with the 
observed \hbox{$L_X - T_X$} relation.  However, we still need a {\it physical 
explanation} for why the relationship between a cluster's gas mass and its 
temperature deviates from its self-similar scaling. 

Expressing the total gas mass within the cluster as \hbox{$M_{gas} = f_{gas} M$}, a 
steepening of the \hbox{$M_{gas} - T_X$} relation can be interpreted as a dependence 
of $f_{gas}$ on cluster mass.  That is, if \hbox{$M \propto T_X^{1.5}$}, as suggested 
by the self-similar model, then the observed \hbox{$M_{gas} - T_X$} relation implies 
that \hbox{$f_{gas} \propto T_X^{0.1 - 0.5}$}.  A varying gas mass fraction is 
expected if the efficiency of galaxy formation varies systematically across clusters 
of different mass.  Observational support for this has been claimed recently by Bryan 
(2000).  However, this is still controversial, and there is no compelling evidence for 
a variation of $f_{gas}$ with cluster temperature (but see Arnaud \& Evrard 1999; 
MME99).  This is especially true for the systems that we are specifically interested 
in: hot clusters with $T_X \gtrsim 3$ keV.  This is apparent, for example, in Figure 
1 (top) of Balogh et al. (2001), who carry out an accounting of stars and gas to 
estimate the fraction of cooling baryons in clusters.  Moreover, Roussel, 
Sadat, \& Blanchard (2000) have carried out a careful analysis of group and cluster 
X-ray data to estimate $f_{gas}$ directly and have found no trends.  More recently, 
Grego et al. (2001) have analysed Sunyaev-Zeldovich effect observations of 18 hot 
clusters and have also found no correlations between a hot cluster's gas mass fraction 
and its temperature.  Finally, observational studies of the {\it total} cluster mass 
($M$) - temperature relation have indicated that $M \propto T_X^{1.6 - 2.0}$ (Horner 
et al. 1999; Ettori \& Fabian 1999; Nevalainen et al. 2000; Finoguenov et al. 2001), 
which, given the observed $M_{gas} - T_X$ relation, is consistent with $f_{gas}$ being 
constant.
 
Theoretically, it is only now becoming possible to reliably investigate the dependence 
of $f_{gas}$ on temperature with the inclusion of radiative cooling, star 
formation, feedback, and other relevant processes in numerical simulations (e.g., 
Lewis et al. 2000; Pearce et al. 2000; Muanwong et al. 2001; Dav\'{e} et al. 2001).  
As of yet, however, there is little agreement in the approaches adopted to model 
these processes and prevent the so-called cooling crisis (compare, for example, the 
findings of Lewis et al. 2000 with those of Pearce et al. 2000).  This is not 
surprising.  As discussed in detail by Balogh et al. (2001), attempting to model the 
effects of cooling across the wide range of halo masses found in clusters is inherently 
very difficult.  The addition of ``sub-grid'' processes, such as star formation and 
feedback, further complicates matters.  Thus, the effects that these additional 
physical processes have on the gas mass fraction of clusters will not be fully 
realized until such issues are resolved.  

In this paper, however, we show that the observed variation of the $M_{gas} 
- T_X$ relation(s) arises quite naturally within the class of models that invoke 
preheating of the intracluster medium during the early stages of cluster formation.  
In these models, $f_{gas}$ is constant on cluster scales ($T_X \gtrsim 3$ keV), and 
the self-similarity is instead broken by an entropy floor generated by early 
non-gravitational heating events.  Preheating has previously been shown to bring 
consistency between a number of other observed and predicted scaling relations for 
groups and clusters (e.g., BBLP02), and therefore one might expect that the $M_{gas} 
- T_X$ relation should also be modified.

The preheating model was originally put forward by Kaiser (1991) and has subsequently 
been investigated by a number of authors (e.g., Evrard \& Henry 1991, Bower 1997, 
Cavaliere et al. 1997; 1998; 1999; Balogh et al. 1999, Wu, Fabian, \& Nulsen 2000; 
Loewenstein 2000, Tozzi \& Norman 2001; Borgani et al. 2001; Thomas et al. 2002; 
BBLP02).  If the ICM is injected with enough thermal energy, the hot X-ray 
emitting gas will become decoupled from the dark halo potential and break the 
self-similar scaling relations.  The best estimates suggest that a substantial amount 
of energy ($\sim$ 1 keV per particle) is required to reproduce the observed relations 
(mainly the $L_X - T_X$ relation).  It is not yet known what source(s) could inject 
such a large amount of energy into the ICM.  Both galactic winds (driven by 
supernovae) and ejecta from active galactic nuclei have been proposed, but 
because of the complexity of the physics, the exact details have yet to be worked 
out.  For an in-depth discussion of potential sources of preheating and of 
alternative possibilities for reproducing the observed relations we refer the reader 
to BBLP02.

In this paper, we adopt the physically motivated, analytic model developed in 
BBLP02 to explore the impact of cluster preheating on the 
$M_{gas} - T_X$ relation.  In comparison with the $L_X - T_X$ and $M - T_X$ 
relations, it has drawn very little attention by theoretical studies.  The 
only studies to have examined the effects of entropy injection on the $M_{gas} - 
T_X$ relation to date are Loewenstein (2000) and Bialek et al. (2001).  To be 
specific, Loewenstein (2000) considered models where the entropy injection occurs at 
the centers of groups and clusters, after the latter have formed whereas Bialek et 
al (2001), like BBLP02, investigated preheated models.  For reasons 
that will be 
described below (in \S 5), we believe our work greatly improves upon both of 
these studies.  The prevailing apathy by theorists is perhaps due in part to a 
near absence of published observational studies on the gas masses of clusters.  
However, in light of the recent important observations discussed above (e.g., 
Vikhlinin et al. 1999; MME99; Neumann \& Arnaud 2001) and the new influx of high 
resolution data from the {\it Chandra} and {\it XMM-Newton} X-ray satellites, which 
will likely 
provide even tighter constraints, we believe that a thorough examination of the $M_{gas} 
- T_X$ relation is timely. 

The models we consider below were developed in a flat $\Lambda$-CDM cosmology with 
$\Omega_m = 0.3$, $h = 0.75$, and a nucleosynthesis value $\Omega_b = 0.019 h^{-2}$ 
(Burles \& Tytler 1998).  They are computed for a number of different preheating 
levels, corresponding to entropy constants of $K_0$ = 100, 200, 300, \& 427 keV 
cm$^2$.  These span the range required to match the observed $L_X - T_X$ relations of 
groups and hot clusters (e.g., Ponman et al. 1999; Lloyd-Davies et al. 2000; Tozzi \& 
Norman 2001; BBLP02).  For the purposes of comparison, we also implement an 
``isothermal'' model (see Section 2.3 in BBLP02), which mimics the self-similar result 
deduced from numerical simulations (e.g., Evrard et al. 1996).

\section{Cluster models}

Since an in-depth discussion of the preheated cluster models can be found 
in BBLP02, we present only a brief description of the models here.

The preheated models can be summarized as follows: the dominant dark matter 
component, which is unaffected by the energy injection, collapses 
and virializes to form bound halos.  The distribution of the dark matter in such 
halos is assumed to be the same as found in recent ultra-high resolution numerical 
simulations (Moore et al. 1998; Klypin et al. 1999; Lewis et al. 2000) and is 
described by
\begin{equation}
\rho_{dm}(r) = \rho_{dm,0} \biggl(\frac{r}{r_s}\biggr)^{-n} \biggl(1 +
\frac{r}{r_s}\biggr)^{n-3}
\end{equation}
\noindent where n = 1.4, $\rho_{dm,0}$ is the profile normalization, and $r_s$ 
is the scale radius. While the dark component is unaffected by energy injection, the 
collapse of the baryonic component is hindered by the pressure forces induced by 
preheating.  If the maximum infall velocity due purely to gravity of the dark halo is 
subsonic, the flow will be strongly affected by the pressure and it will not undergo 
accretion shocks.  It is assumed that the baryons will accumulate onto the halos {\it 
isentropically} at the adiabatic Bondi accretion rate (as described in Balogh et al. 
1999).  This treatment, however, is only appropriate for low mass halos.  If the 
gravity of the dark halos is strong enough (as it is expected to be in the hot 
clusters being considered here) so that the maximum infall velocity is 
transonic or supersonic, the gas will experience an additional (generally dominant) 
entropy increase due to accretion shocks.  In order to trace the shock history of the 
gas, a detailed knowledge of the merger history of the cluster/group is required but is 
not considered by BBLP02.  Instead, it is assumed that at some earlier time 
the most massive cluster progenitor will have had a mass low enough such that shocks 
were negligible in its formation, similar to the low mass halos discussed above.  This 
progenitor forms an isentropic core of radius $r_{c}$ at the cluster center.  The 
entropy of gas outside of the core, however, will be affected by shocks.  Recent high 
resolution numerical simulations suggest that the entropy profile for gas outside this 
core can be adequately represented by a simple analytic expression given by $\ln{K(r)} 
= \ln{K_0} + \alpha \ln{(r/r_c)}$ (Lewis et al. 2000), where $\alpha \sim 1.1$ for 
the massive, hot clusters ($T_X \gtrsim 3$ keV) of interest here (Tozzi \& Norman 2001; 
BBLP02).  It should be noted that in the case of these massive systems, the accretion 
of gas is limited by gravitational infall, and hence they accrete their full 
compliment of baryons [i.e., $M_{gas} = (\Omega_b / \Omega_m) M$].  It is 
assumed that the mass of baryons locked up in stars is negligible (as suggested by, 
for example, Roussel et al. 2000; Balogh et al. 2001). 

Following this prescription and specifying the parameters $r_c$, $\rho_{gas}(r_c)$, 
and $\alpha$ (as discussed in BBLP02) completely determines the models.  Under all 
conditions, the gas is assumed to be in hydrostatic equilibrium within the dark 
halo potential.  The effects of radiative cooling are neglected by these models.

\section{How Preheating Affects the \hbox{$M_{gas} - T_X$} Relation}

Preheating will affect the $M_{gas} - T_X$ relation in two ways: (1) by altering 
the temperature profile and increasing the emission-weighted gas temperature of a 
cluster and (2) by altering the gas density profile and reducing the gas mass in the 
cluster core.  We are interested in the strength of these effects and whether or not 
they can be distinguished by current or future observational data.  First, we consider 
the effect of preheating on the temperature of a cluster.

Figure 1 is a plot of $T_X$ as a function of entropy floor level (i.e., 
$K_0$) for three clusters of different {\it total} masses.  The thin line represents a 
cluster with $M \approx 5.6 \times 10^{14} M_{\odot}$, the next thickest line 
represents a cluster with $M \approx 10^{15} M_{\odot}$, and the thickest line 
represents a cluster with $M \approx 1.8 \times 10^{15} M_{\odot}$.  As expected, the 
gas temperature of a cluster increases as the level of preheating is increased.  On 
average, an increase in $T_X$ of about 1 keV (10 - 25\%) occurs when a cluster is 
preheated to the level of $K_0 \gtrsim 300$ keV cm$^2$ (over the range 3 keV $\lesssim 
T_X \lesssim 10$ keV).  This effect will primarily manifest itself as a normalization 
shift in the $M_{gas} - T_X$ relation.  

\placefigure{fig1}
\placefigure{fig2}

Figure 2 presents the dimensionless gas density profile of a cluster with $T_X = 4$ keV 
(left panel) and a cluster with $T_X = 8$ keV (right panel) as a function of the level 
of preheating.  The dot-dashed line is the self-similar result (i.e., isothermal model 
of BBLP02).  The long-dashed, short-dashed, dotted, and solid lines represent the 
preheated models of BBLP02 with $K_0$ = 100, 200, 300, and 427 keV cm$^2$, 
respectively.  
Preheating reduces the gas density, and therefore the gas mass, in the central 
regions of a cluster.  In \S 4, we investigate the $M_{gas} - T_X$ relation within 
three different radii: $r = 0.25 h^{-1}$ Mpc and $0.50 h^{-1}$ Mpc and $r_{500}$  
(the radius within which the mean dark matter mass density is 500 times the mean 
critical density $\rho_{crit}$ at $z$ = 0).  These radii are indicated in Figure 2 by 
the open squares, pentagons, and triangles, respectively.  Clearly, the effect on the 
$M_{gas} - T_X$ relation will be strongest when $M_{gas}$ is evaluated within $r = 0.25 
h^{-1}$ Mpc.  
{\epsscale{1.0} \plotone{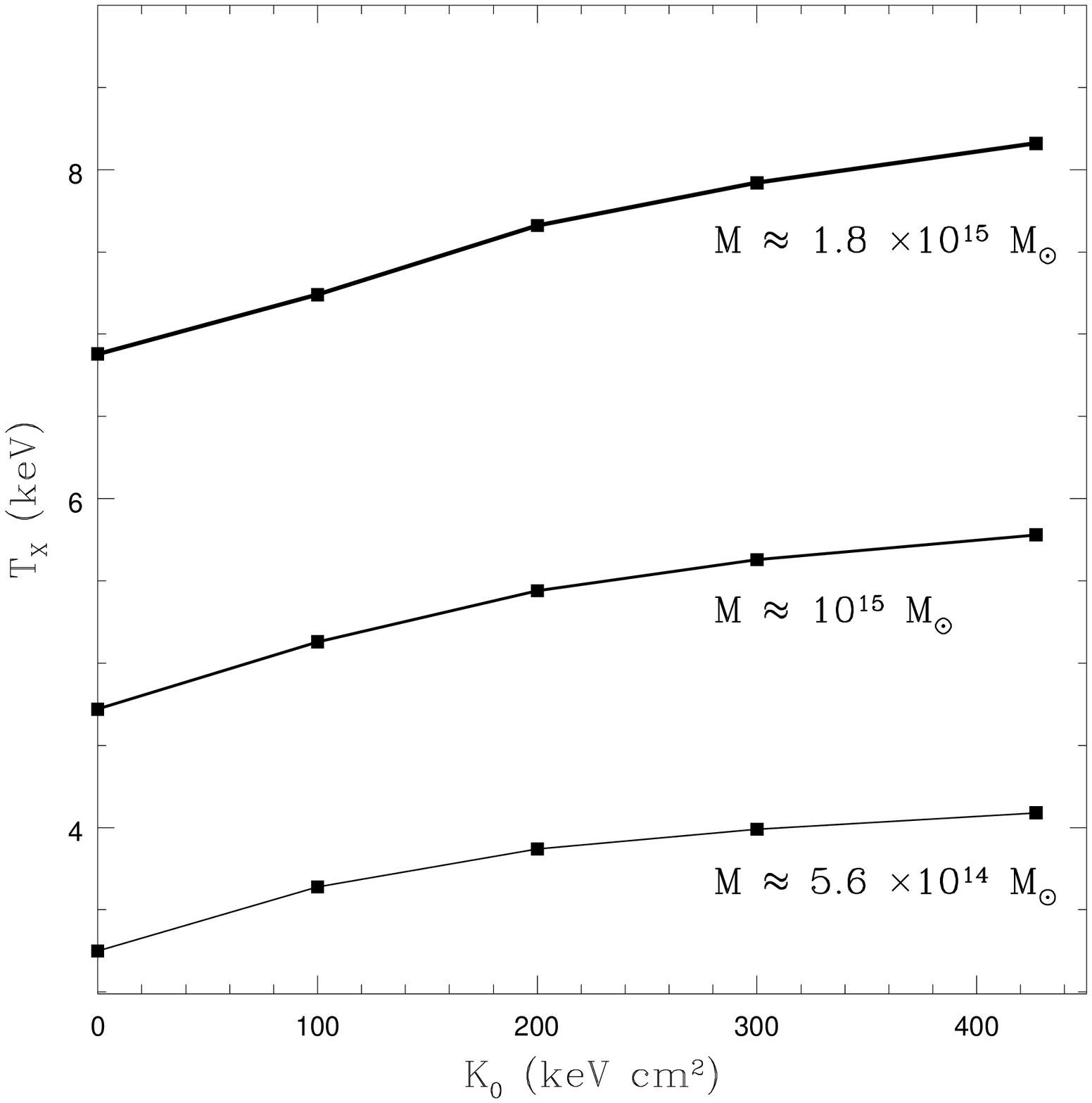}
{Fig.1. \footnotesize Effect of preheating on a cluster's temperature.  The thin
line, the next thickest line, and the thickest line represent clusters with $M 
\approx
5.6 \times 10^{14}$, $10^{15}$, and $1.8 \times 10^{15} M_{\odot}$, respectively.  
The
solid squares indicate the discrete points where the model was actually evaluated.}}
\vskip0.1in
\noindent
Furthermore, because $r = 0.25 h^{-1}$ Mpc is a fixed radius that samples 
different fractions of the virial radius ($R_{halo}$) for clusters of different 
temperature, the (fractional) reduction in gas mass within that radius 
{\epsscale{1.0} \plotone{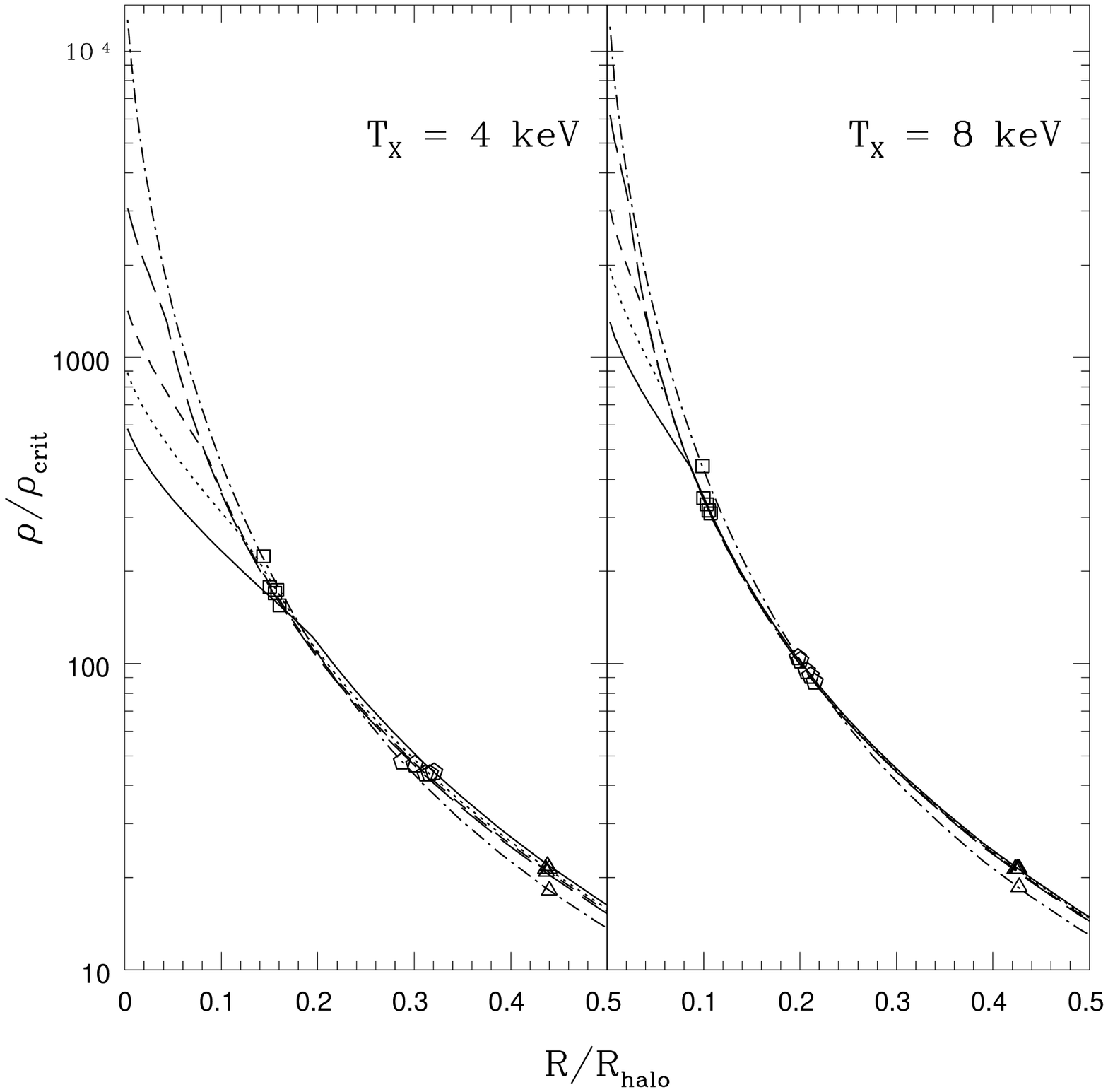}
{Fig.2. \footnotesize Effect of preheating on a cluster's gas density profile.
{\it Left}: Cluster with $T_X = 4$ keV. {\it Right}: Cluster with $T_X = 8$ keV.  The
dot-dashed line is the self-similar result.  The long-dashed, short-dashed, dotted,
and solid lines represent the preheated models of BBLP with $K_0$ = 100, 200, 300,
and
427 keV cm$^2$, respectively.  The squares, pentagons, and triangles indicate the
radii $r = 0.25 h^{-1}$ Mpc and $0.50 h^{-1}$ Mpc and $r_{500}$, respectively, for
each of the models.}}
\vskip0.1in
\noindent
will be largest for the lowest temperature systems.  This will lead to both 
a normalization shift and a steepening of the $M_{gas} - T_X$ relation.  To 
illustrate how strong the effect is, we 
examine the reduction of gas mass within $r = 0.25 h^{-1}$ Mpc, $0.50 h^{-1}$ Mpc and 
$r_{500}$ for a low mass 
cluster with a {\it total} mass of $M \approx 5.6 \times 
10^{14} M_{\odot}$ and for a high mass cluster with $M \approx 1.8 \times 10^{15} 
M_{\odot}$.  We find that when the low mass cluster has undergone preheating at the 
level of $K_0 = 100$ keV cm$^2$ it has $\approx 32\%$ more mass in gas within $r = 
0.25 h^{-1}$ Mpc than when the same cluster has undergone preheating at the level of 
$K_0 = 427$ keV cm$^2$.  Using the same test on the $M \approx 1.8 \times 10^{15} 
M_{\odot}$ cluster, however, yields a difference of only 22\%.  When we probe the 
larger radius $r = 0.50 h^{-1}$ Mpc, we find the effect is less pronounced (as 
expected).  The difference in $M_{gas}$ between the $K_0 = 100$ keV cm$^2$ model and 
$K_0 = 427$ keV cm$^2$ model is 8\% for the low mass cluster and 7\% for 
the high mass cluster.  Finally, when the gas mass is evaluated within $r_{500}$, the 
difference is 4\% for the low mass cluster as opposed to 2\% for the high mass 
cluster.

In summary, preheating will significantly affect the $M_{gas} - T_X$ relation by 
increasing the emission-weighted gas temperature of clusters.  Whether or not the 
relation is also affected by the reduction of gas mass in the cores of clusters depends 
on within which radius $M_{gas}$ is evaluated and what temperature regime is being 
probed.  The effect will be strongest for low temperature systems and when $M_{gas}$ is 
probed within small radii (e.g., $r = 0.25 h^{-1}$ Mpc).  An evaluation of the 
$M_{gas} 
- T_X$ relation within large radii, such as $r_{500}$, however, probes the integrated 
properties of a cluster and will be sensitive only to the temperature shift.  

In the next section, we compare the results of the BBLP02 preheated models to genuine 
observational data.  As we show below, only models with $K_0 \gtrsim 300$ keV cm$^2$ 
are consistent with the data.

\section{Results}

%
%
%
%

In Figure 3 we present the $M_{gas} - T_X$ relation as predicted by the BBLP02 
preheated models within $r_{500}$.  The radius $r_{500}$ is typically comparable in 
size to the {\it observed} radius of a cluster and represents the boundary between 
the inner, virialized region and recently accreted, still settling outer region of a 
cluster (Evrard et al. 1996).  Thus, as already mentioned, the $M_{gas}(r_{500}) - 
T_X$ relation can be regarded as a probe of the integrated properties of a cluster 
and can be directly compared with the self-similar result of $M_{gas} \propto 
T_X^{1.5}$.  

In Figures 4 and 5 we present the $M_{gas} - T_X$ relation as predicted by the BBLP02 
preheated models within the fixed radii $r = 0.25 h^{-1}$ Mpc and $r = 0.50 h^{-1}$ 
Mpc, respectively.  As mentioned above, the determination of the $M_{gas} - T_X$ 
relation within some fixed radius, such as $r = 0.25 h^{-1}$ Mpc or $r = 0.50 
h^{-1}$ Mpc, can be used as an indirect probe of the gas density profiles of 
clusters because it samples different fractions of the virial radius for clusters of 
different temperature.  For the purposes of clarity, we discuss the $M_{gas} - T_X$ 
relation at these three radii separately.

{\epsscale{1.0} \plotone{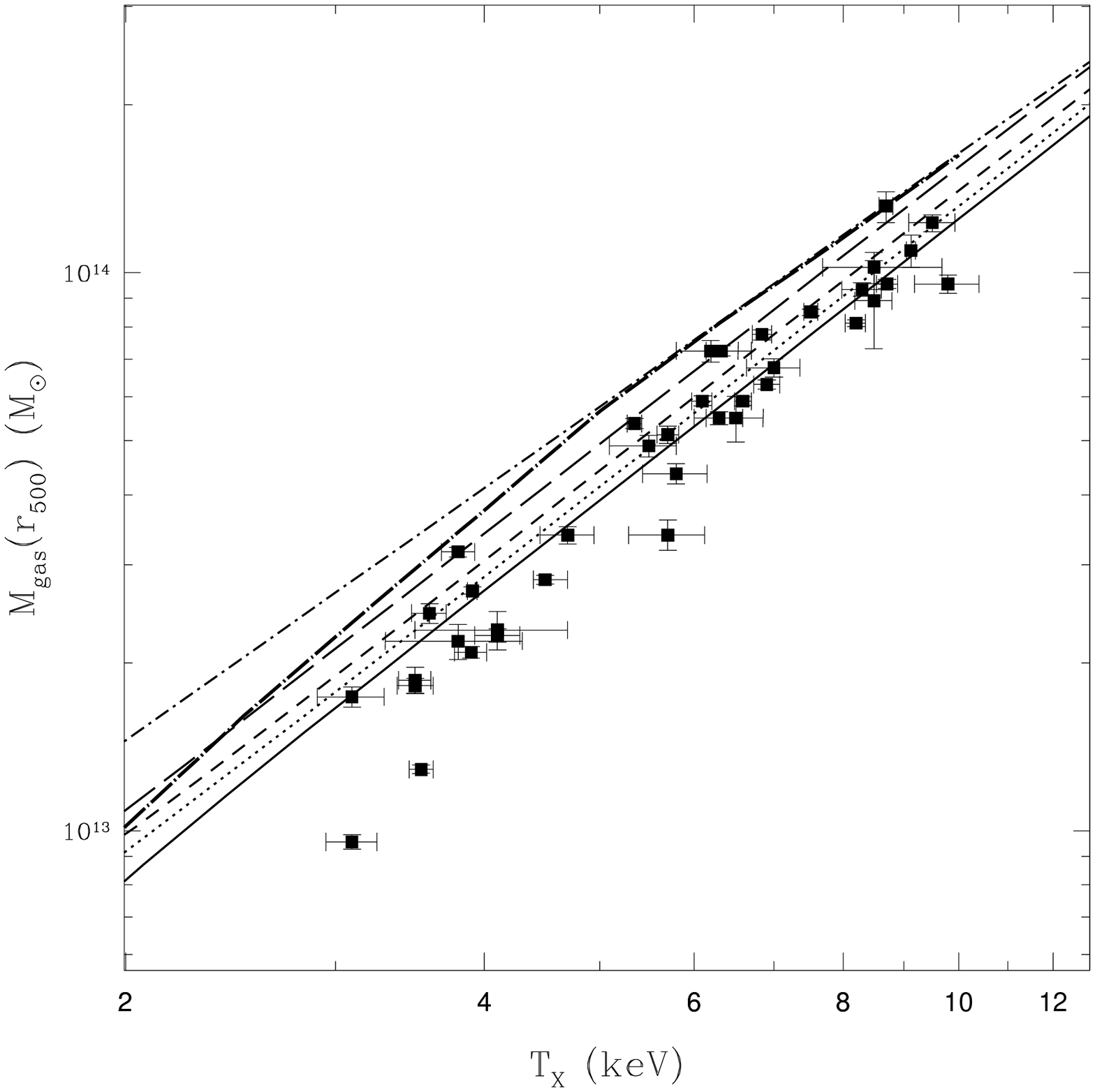}
{Fig.3. \footnotesize Comparison of $M_{gas}(r_{500}) - T_X$ relations.  Squares
represent the gas mass determinations of Mohr et al. (1999) within $r_{500}$.
The dot-short-dashed, long-dashed, short-dashed, dotted, and solid lines represent
the self-similar model and preheated models of BBLP02 with $K_0$ = 100, 200, 300,
and 427 keV cm$^2$, respectively.  The dot-long-dashed (thick) line represents 
the best-fit heated model of Loewenstein (2000).}}

{\epsscale{1.0} \plotone{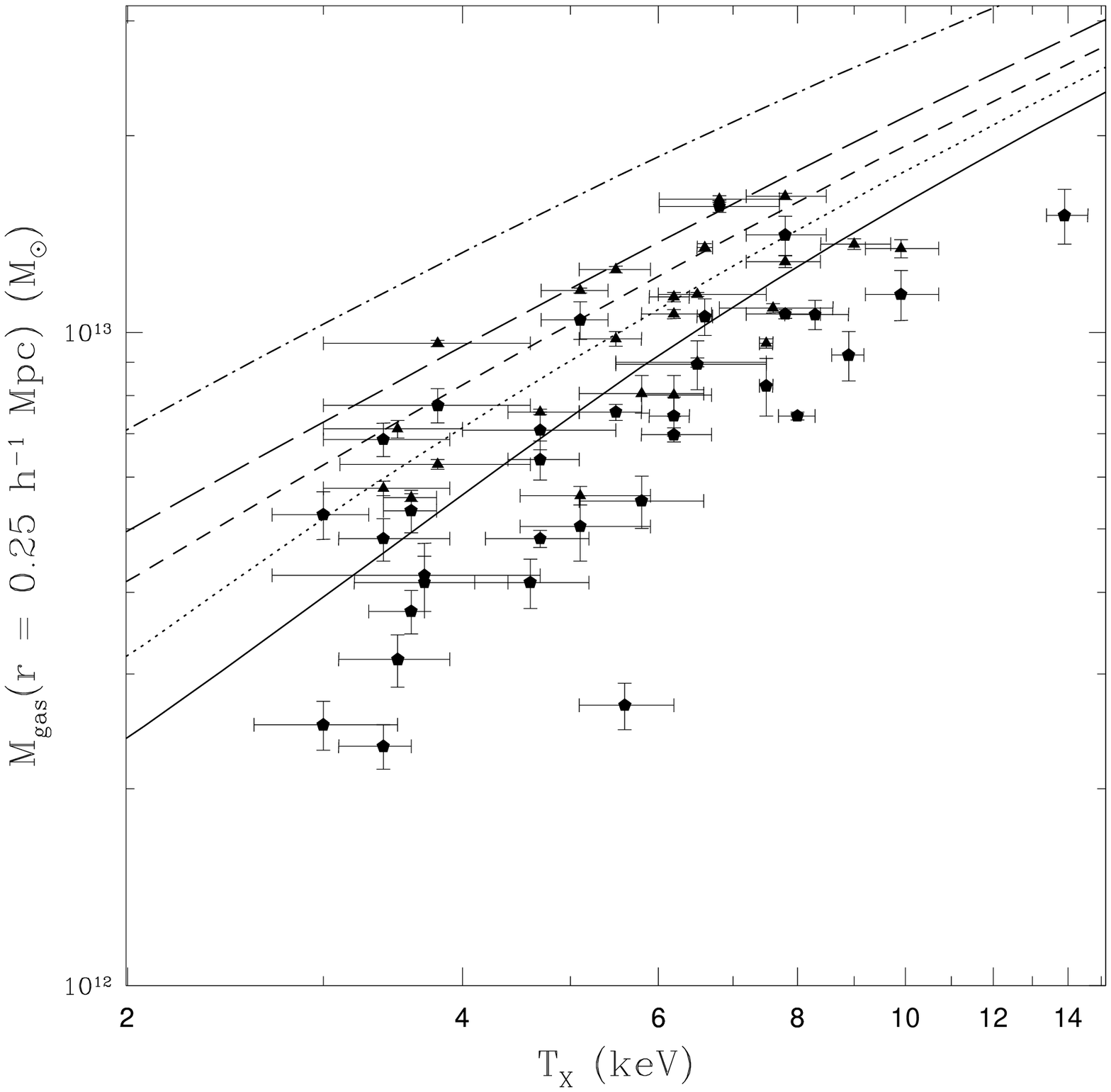}
{Fig.4. \footnotesize Comparison of $M_{gas}$($r = 0.25 h^{-1}$ Mpc)$ - T_X$
relations.  The solid triangles and pentagons represent the gas mass determinations
Peres et al. (1998) and White et al. (1997) within $r = 0.25 h^{-1}$ Mpc,
respectively. The dot-dashed line is the self-similar result.  The long-dashed,
short-dashed, dotted, and solid lines represent the preheated models of BBLP02 with
$K_0$ = 100, 200, 300, and 427 keV cm$^2$, respectively.}}

{\epsscale{1.0} \plotone{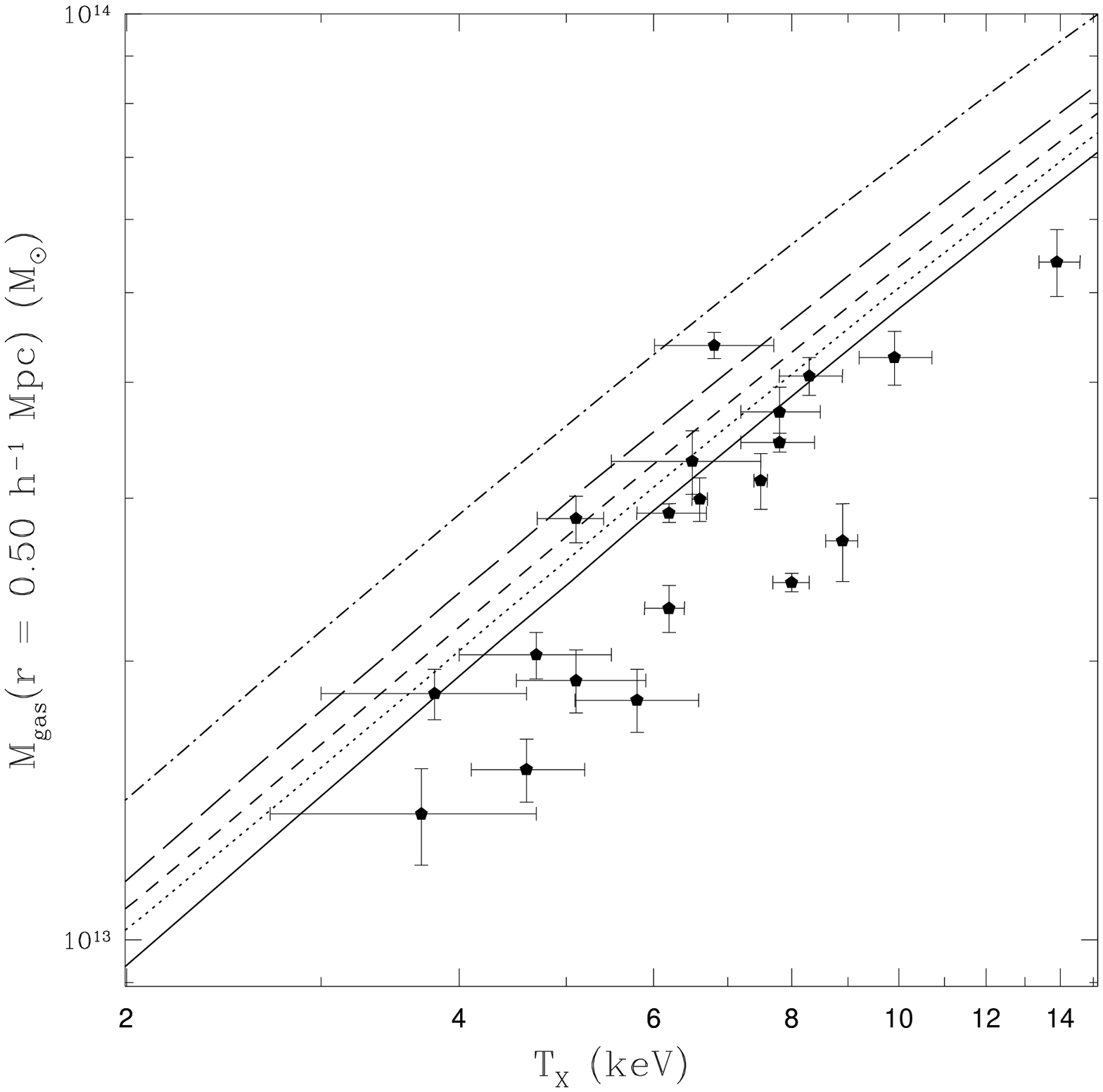}
{Fig.5. \footnotesize Comparison of $M_{gas}$($r = 0.50 h^{-1}$ Mpc)$ - T_X$
relations.  The
solid pentagons represent the gas mass determinations of White et al. (1997)
within $r = 0.50 h^{-1}$ Mpc.  The dot-dashed line is the self-similar result.  The
long-dashed, short-dashed, dotted, and solid lines represent the preheated models of
BBLP02 with $K_0$ = 100, 200, 300, and 427 keV cm$^2$, respectively.}}

\subsection{Test 1: $M_{gas}(r_{500}) - T_X$}

The solid squares in Figure 3 represent the gas mass determinations of MME99 within 
$r_{500}$ using surface brightness profile fitting 
(with isothermal $\beta$ models) of {\it ROSAT} Position Sensitive Proportional 
Counter data and mean emission-weighted 
temperatures from the literature, for clusters with $T_X \geq 3$ keV and whose error 
bars are 1 keV or smaller.  We compare this with the self-similar result represented 
by the ``isothermal'' model of BBLP02 (dot-short-dashed line).  Finally, the 
long-dashed, 
short-dashed, dotted, and solid lines represent the preheated models of BBLP02 with 
$K_0$ = 100, 200, 300, and 427 keV cm$^2$, respectively.  The thick 
dot-long-dashed line represents the predictions of the best-fit heated model of 
Loewenstein (2000).  This model is discussed further in \S 5.1.

\placefigure{fig3}

It is readily apparent that only the preheated models of BBLP02 with $K_0 \gtrsim 200$ 
keV cm$^2$ have a reasonable chance of being consistent with the data of MME99.  The 
normalization clearly indicates that the observed gas temperature of clusters with a 
given gas mass is hotter than predicted by models with entropy floors of $K_0 \lesssim 
100$ keV cm$^2$.  We note that this discrepancy may be remedied by assuming a smaller 
value of $\Omega_b/\Omega_m$.  However, a similar offset, in the same sense, is seen 
in the correlation with {\it total} dark matter mass and gas temperature (Horner et 
al. 1999; Nevalainen et al. 2000; Finoguenov et al. 2001).  This will not be 
reconciled by lowering $\Omega_b/\Omega_m$.  The reason why the preheated models with 
$K_0 \gtrsim 200$ keV cm$^2$ are better able to match the normalization of the 
observational data than models with $K_0 \lesssim 100$ keV cm$^2$ is, as 
mentioned above, because an increase in the amount of preheating directly leads to an 
increase in the emission-weighted gas temperature. 

We have attempted to quantify how well (or poorly) the preheated and self-similar 
models match the observational data.  We have fit both the theoretical results and the 
observational data with simple linear models of the form $\log{M_{gas}} = m \log{T_X} 
+ b$ over the range 3 keV $\lesssim T_X \lesssim 10$ keV.  For the theoretical 
results, we calculate the best-fit slope and intercept using the ordinary least 
squares 
(OLS) test.  We stress that the results of these fits, which are presented in Table 1, 
are only valid for clusters with $T_X \gtrsim 3$ keV.  At lower temperatures, the role 
of preheating becomes much more important [as $M_{gas}$ becomes less than 
$(\Omega_b / \Omega_m) M$] and as a result, the relations steepen 
dramatically.  For example, the preheated model with $K_0 = 427$ keV cm$^2$ is well 
approximated by a power-law with $M_{gas} \propto T_X^{1.68}$ over the range 3 keV  
$\lesssim T_X \lesssim 10$ keV but is significantly steeper over the range 1 
keV $\lesssim T_X \lesssim 3$ keV with $M_{gas} \propto T_X^{1.94}$.  {\it Thus, it 
is absolutely essential that comparisons between theoretical models and observations 
are done over the same range in temperatures.}  

To fit the observational data of MME99, we have used a 
linear regression technique that takes into account measurement errors in both 
coordinates as well as intrinsic scatter (the BCES test of Akritas \& Bershady 1996).  
As a consistency check, we have also employed 10,000 Monte Carlo bootstrap 
simulations.  No significant deviations between the two tests were found.  The results 
of the linear regression fits to the observational data are also presented in Table 1.

\placetable{tab1}

\footnotetext[3]{Our best fit differs slightly from MME99's best fit to their own
data because we implemented a different selection criteria. Namely, we have used only
clusters with $T_X \geq 3$ keV and whose error bars are 1 keV or smaller.}

For all 38 clusters taken from MME99, we derive a best fit that is inconsistent with 
the 
results of {\it all} of the theoretical models considered at greater than the 90\% 
confidence level$^1$.  However, as is apparent from Figure 3, the slope and 
intercept of the best-fit line are sure to be heavily dependent upon the two low 
temperature clusters with the lowest measured gas masses (and gas mass fractions): the 
Hya I cluster (Abell 1060) and the Cen cluster (Abell 3526).  A number of other 
studies (both optical and X-ray) have also identified very unusual properties in both 
clusters.  For example, Fitchett \& Merritt (1988) were unable to fit a spherical 
equilibrium model to the kinematics of galaxies in the core of Hya I.  They suggest 
that substructure is present and is likely why Hya I does not lie 
along the $L_X - \sigma$ relation for galaxy clusters.  More recently, Furusho et al. 
(2001) have found that the metal abundance distribution implies that the gas in Hya 
I is well-mixed (i.e., it does not contain an obvious metallicity gradient), 
suggesting 
that a major merger event may have occurred sometime after the enrichment of the ICM. 
 Measurements of the bulk motions of the intracluster gas in the Cen cluster 
(through Doppler shifting of X-ray spectral lines) reveal strange gas velocity 
gradients indicative of a large merger event in the not too distant past (Dupke \& 
Bregman 2001).  This picture has also been supported by Furusho et al. (2001), who 
found large temperature variations across the cluster's surface.  Thus, neither Hya 
I nor Cen can be regarded as typical ``relaxed'' clusters and are probably not 
representative of the majority of low temperature systems.  

One way to ameliorate the impact of the two clusters would be to increase the number of 
systems of this temperature.  However, there are very few published gas 
mass estimates of cool clusters and groups within $r_{500}$.  X-ray emission from 
groups is usually only detected out to a small fraction of this radius.  The one study 
that does present group gas masses for a radius at fixed overdensity, Roussel et al. 
(2000), does so for $r_{200}$ and is not directly comparable to the results presented in 
Figure 3.  Also, in that study, gas masses were determined by extrapolating the surface 
brightnesses far outside the limiting radius for which X-ray emission was actually 
detected.  This can lead to biases in determining group/cluster properties (see 
Mulchaey 2000; Balogh et al. 2001).

In recognition of the above, we have tried removing Hya I and Cen from 
the sample and fitting the remaining 36 clusters using the same procedure.  We find 
that the preheated model with $K_0 = 427$ keV cm$^2$ is then consistent with the data 
at the 90\% level.  The $K_0$ = 200 and 300 keV cm$^2$ models are marginally 
inconsistent with the MME99 data.  The isothermal model is ruled out at $\gtrsim  
99\%$ 
confidence irrespective of whether these clusters are dropped or not.

The other two observational studies that have investigated the $M_{gas} - T_X$ 
relation, Neumann \& Arnaud (2001) and Vikhlinin et al. (1999), unfortunately did not 
present gas mass determinations within $r_{500}$ for individual clusters in a table or 
graphically.  They did, however, present their best-fit values for the slope of the 
relation.  These were deduced from samples of clusters that have temperatures 
spanning roughly the same range as that considered in Figure 3.  The best-fit slopes 
of the preheated models are shallower than the best-fit claimed by Neumann \& 
Arnaud (2001) of $M_{gas} \propto T_X^{1.94}$ for a sample of 15 hot clusters.  
However, an estimate of the uncertainty on this result was not reported; thus we are 
unable to say whether this result is inconsistent with the predictions of the 
preheated models.  The predicted slopes of all four preheated models studied here are 
in excellent agreement with the findings of Vikhlinin et al. (1999), who report 
$M_{gas} \propto T_X^{1.71 \pm 0.13}$ for their sample of 39 clusters.  We also note 
that the results of Neumann \& Arnaud (2001) and Vikhlinin et al. (1999) differ 
significantly from the predictions of the self-similar model.

In summary, we find that the class of models that invoke preheating are much 
better 
able to match the observed $M_{gas}(r_{500}) - T_X$ of hot clusters than that of the 
isothermal self-similar model, which is ruled out with a high level of 
confidence.  A careful analysis of the MME99 data also suggests that only those models 
that invoke a ``high'' level of energy injection (i.e., $K_0 > 300$ keV cm$^2$) are 
able to match observations. 

\subsection{Test 2: $M_{gas}$($r = 0.25 h^{-1}$ Mpc) - $T_X$}

The solid triangles and pentagons in Figure 4 represent the gas mass determinations 
within $r = 0.25 h^{-1}$ Mpc of Peres et al. (1998) and White et al. (1997), 
respectively.  These data were obtained using surface brightness profile fitting of 
{\it ROSAT} data (Peres et al. 1998) and {\it Einstein} data (White et al. 1997) and 
emission-weighted temperatures from the literature, for clusters with $T_X \geq 3$ keV 
and whose error bars are 1 keV or smaller.  The predictions of the isothermal 
self-similar model are represented by the dot-dashed line.  Once again, the 
long-dashed, short-dashed, dotted, and solid lines represent the preheated models of 
BBLP02 with $K_0$ = 100, 200, 300, and 427 keV cm$^2$, respectively.

\placefigure{fig4}

In spite of the scatter, it is apparent that only those preheated models with entropy 
floors of $K_0 \gtrsim 300$ keV cm$^2$ are consistent with the 57 clusters plotted in
Figure 4.  As with the $M_{gas}(r_{500}) - T_X$ relation, the normalization of the
self-similar model and the preheated model with $K_0 = 100$ keV cm$^2$ suggests that
ICM is observed to be much hotter than predicted by either of these models.  Fitting
both the theoretical predictions and observational data in a manner identical to 
that presented in the previous subsection, we find that only the preheated models 
with $K_0 \geq 300$ keV cm$^2$ have both slopes and intercepts that are consistent 
with the observational data (see Table 2).  On the basis of normalization (intercept), 
the self-similar model is ruled out with greater than 99\% confidence.  

In \S 3 we briefly discussed the potential of the gas density profile to affect
the $M_{gas}$($r = 0.25 h^{-1}$ Mpc) - $T_X$ relation.  This effect is obvious in
Figure 4, with mild breaks at $T_X \approx 10$ keV for the $K_0 = 427$ keV cm$^2$ model 
and at $T_X \approx 5$ keV for the $K_0 = 300$ keV cm$^2$ model.  However, with the 
large scatter obscuring any potential breaks in the $M_{gas} - T_X$ relation, all we 
can conclude is that the data are consistent with predicted profiles of the BBLP02 
preheated models with $K_0 \gtrsim 300$ keV cm$^2$.

\placetable{tab2}

The exact nature of the scatter in Figure 4 is unclear.  While some of the scatter is 
likely attributable to the large uncertainties in the temperature measurements made 
using {\it Einstein}, {\it Ginga}, and {\it EXOSAT} data, some of it may also be due 
to unresolved 
substructure (e.g., cooling flows) and  point sources.  Such issues become 
particularly important when investigating the central regions of clusters as opposed 
to its integrated properties.  Indeed, new high-resolution data obtained by 
{\it Chandra} support this idea (see, e.g., Stanford et al. 2001).  We anticipate that 
future data obtained by both {\it Chandra} and {\it XMM-Newton} will place much 
tighter 
constraints on the $M_{gas}$($r = 0.25 h^{-1}$ Mpc) - $T_X$ relation and possibly 
even allow one to probe the mild break in the relationship predicted by the preheated 
models.

\subsection{Test 3: $M_{gas}$($r = 0.50 h^{-1}$ Mpc) - $T_X$}

The solid pentagons in Figure 5 represent the gas mass determinations
of White et al. (1997) within $r = 0.50 h^{-1}$ Mpc using surface profile 
fitting of {\it Einstein} data and emission-weighted gas temperatures from the 
literature, 
for clusters with $T_X \geq 3$ keV and whose error bars are 1 keV or smaller.  Again, 
the predictions of the isothermal self-similar model are represented by the
dot-dashed line while the long-dashed, short-dashed, dotted, and solid
lines represent the preheated models with $K_0$ = 100, 200, 300, and 427 keV
cm$^2$, respectively.

\placefigure{fig5}
\placetable{tab3}

A linear regression fit to the 20 clusters taken from White et al. (1997) 
yields a best-fit slope and intercept that is consistent with only the $K_0$ = 300 
and 427 keV cm$^2$ preheated models (90\% confidence; see Table 3).  Once again, the 
isothermal model is ruled out with greater than 99\% confidence.  This follows the 
same general trend discovered in the previous tests.

As expected, the influence of the gas density profile on the $M_{gas}$($r = 0.50 
h^{-1}$ Mpc) - $T_X$ is minimal.  Only a very modest break is detectable at $T_X 
\approx 4.5$ keV for the $K_0 = 427$ keV cm$^2$ preheated model.  Like the 
$M_{gas}(r_{500}) - T_X$ relation, this relation is mostly sensitive to the temperature 
shift (at least for clusters with $T_X \gtrsim 3$ keV).

\section{Comparison with previous theoretical studies}

Only two other theoretical studies have examined the effects on the $M_{gas} - 
T_X$ relation of entropy injection into the ICM: Loewenstein (2000) and Bialek et 
al. (2001).  Both studies investigated the $M_{gas}(r_{500}) - T_X$ relation and 
demonstrated that entropy injection does, indeed, steepen the relation, in 
agreement with the present work (however, neither implemented the $M_{gas} - T_X$ at 
fixed radii test).  These studies suggest that models that produce an 
entropy floor with a level that is consistent with measurements of groups ($K_0 \sim 
100$ keV cm$^2$; Ponman et al. 1999; Lloyd-Davies et al. 2000) are capable of 
matching the observations of even hot clusters (up to 10 
keV).  This is in apparent conflict with the results presented in \S 3 that 
suggest that a high entropy floor of $K_0 \gtrsim 300$ keV cm$^2$ is required to match 
the observations of hot clusters.  A low value of the entropy floor is also in 
apparent conflict with a number of other studies that have focused mainly on the $L_X 
- T_X$ relation of hot clusters.  For example, da Silva et al. (2001), Tozzi \& Norman 
(2001) and BBLP02 have all concluded that such low levels of entropy injection do not 
bring consistency between observations and theoretical models of {\it hot clusters}.  
As such, a closer analysis of Loewenstein (2000) and Bialek et al. (2001) studies is 
warranted.    

\subsection{The Loewenstein (2000) Models}

To model the observed deviations of the cluster X-ray scaling relations, Loewenstein 
(2000) has constructed a suite of hydrostatic polytropic models (which are normalized 
to observations of high-temperature clusters and numerical simulations), and then 
modified them by adding various amounts of heat per particle at the cluster center.  
Strictly speaking, the Loewenstein (2000) models cannot be characterized as {\it 
preheated} models, since the injection of entropy into the ICM occurs after the 
cluster has formed.  Thus, a straightforward comparison between the Loewenstein (2000) 
and BBLP02 models is not trivial.  However, success in matching the $M_{gas}(r_{500}) 
- T_X$ relation (the data of MME99) is claimed by Loewenstein (2000) for a model that 
``produces an entropy - temperature relation with the observed entropy floor at 
$\approx 100$ keV cm$^2$.''  Regardless of how the entropy floor actually arose, this 
contradicts the results presented in \S 3, which suggest that an entropy floor of 
$\gtrsim 300$ keV cm$^2$ is required to match the observations.  Can the analysis of 
Loewenstein (2000) and that of the present work be reconciled?

A closer investigation of Figure 4 of Loewenstein (2000) reveals that first of all, 
his heated models were not compared to the actual data but rather to points that 
represent MME99's best-fit power-law match to their data.  Second, this power-law 
relationship was assumed to hold true and hence, extrapolated to span a wider range 
in temperatures than considered by MME99.  Of the 45 clusters studied by MME999, 
only one had a temperature below 3 keV (it was 2.41 keV), yet Loewenstein (2000) 
compared his heated models to the best-fit relation of MME99 over the range 1 keV 
$\lesssim T_X \lesssim 10$ keV.  As previously mentioned, however, entropy injection 
preferentially affects low-temperature systems and, therefore, extrapolating scaling 
relations derived from high-temperature systems down to the low-temperature regime is 
not safe.

In Figure 3, we compare the best-fit heated model of Loewenstein (2000) (his 
$\epsilon = 0.35$ model, as the thick dot-long-dashed line) with the predictions of 
BBLP02 models and 
the data of MME99.  The plot clearly demonstrates that his best-fit model does not 
match the data of MME99 nearly as well as the BBLP02 preheated models with $K_0 
\gtrsim 300$ keV cm$^2$, especially at the high temperature end.  The difference in 
temperature ranges examined by Loewenstein (2000) and the present study (whose range 
of temperatures were purposely chosen to match the observational data) has likely 
led to an underestimation of the entropy floor in these clusters by Loewenstein 
(2000).  We once again re-iterate that it is extremely important that comparisons 
between theoretical models and observations are done over the same range in 
temperatures.

\subsection{The Bialek et al. (2001) simulations}    

In similarity to the present work, Bialek et al. (2001) investigated the impact of 
preheating on the $M_{gas} - T_X$ relation for a number of different levels of entropy 
injection, spanning the range 0 keV cm$^2$ $\lesssim K_0 \lesssim 335$ keV cm$^2$.  
Fitting their 
$M_{gas}(r_{500}) - T_X$ simulation data over the range 2 keV $\lesssim T_X \lesssim 
9$ keV, which is similar (but not identical) to the MME99 sample, they claim 
success in matching the observations of MME99 for models with entropy injection at 
the level of 55 keV cm$^2$ $\lesssim K_0 \lesssim 140$ keV cm$^2$, at least on the 
basis of slope.  Their models with higher levels of entropy injection, apparently, 
predict relations much too steep to be consistent with the data of MME99.  These 
predictions are inconsistent with the results of the BBLP02 analytic models with 
similar levels of entropy injection (e.g., for $K_0 \approx 300$ keV cm$^2$, BBLP02 
predict $M_{gas} \propto T_X^{1.67}$ while Bialek et al. find $M_{gas} \propto 
T_X^{2.67}$) .  However, we believe the difference in the predictions (and 
conclusions) of Bialek et al. (2001) and the present work can be reconciled.  

As noted by Neumann \& Arnaud (2001), Bialek et al. (2001) have simulated very few hot 
clusters and, although they fit their $M_{gas}(r_{500}) - T_X$ simulation data over a 
range similar to MME99, the results are too heavily weighted by the cool clusters 
($T_X \lesssim 3$ keV) to be properly compared with the data of MME99.  As an 
example, we consider their ``S6'' sample of 12 clusters that have $K_0 = 335$ 
keV cm$^2$.  According to the present study, this model should give a reasonably good 
fit to the MME99 observational data, much better than that of a model with $K_0 
\approx 100$ keV cm$^2$.  Although the normalization of the S6 model is in excellent 
agreement with the MME99 data (as is apparent in Table 3 of Bialek et al. and the 
general trends in their Figure 1) they rule this model out based on the fact that the 
predicted slope is 2.67, much steeper than the 1.98 found by MME99.  However, a 
closer analysis reveals that the fraction of cool clusters in the simulation data set 
is much higher than fraction of cool clusters in the MME99 sample.  For example, in 
the MME99 sample of 45 clusters, only one cluster has a temperature below 3 keV.  In 
the Bialek et al. (2001) S6 set, however, 5 of the 12 clusters have temperatures 
below 3 keV.  In addition, the mean temperature of clusters in the MME99 sample is 
$\approx$ 5.5 keV, while it is only about 3.8 keV in the Bialek et al. (2001) S6 data 
set.  As previously mentioned, preheating preferentially affects low temperature 
systems and, therefore comparisons between theory and observations should be done 
over the same range in temperatures.  To illustrate the problems of comparing 
theoretical models and observations that span different temperature ranges, we tried 
to reproduce the fit of Bialek et al. (2001) to their S6 data set.   We used 
data presented in their Table 2 for clusters with $T_X > 2$ keV (we use their 
preferred ``processed'' temperatures) and fit it with a linear model and found $M_{gas} 
\propto T_X^{2.42 \pm 0.17}$.  This is slightly different from the value listed in 
their Table 3, presumably because Table 2 is based on data within $r_{200}$ while Table 
3 is based on data within $r_{500}$ (they note that a change of up to 6\% in the 
predicted slope can occur when switching between the two).  To match the conditions of 
the present work, we then discarded all simulated cluster data below 3 keV (the mean 
temperature for the remaining 7 clusters was then $5.1$ keV, similar to the MME99 
data) and found a best fit of $M_{gas} \propto T_X^{1.99 \pm 0.30}$.  This is 
excellent agreement with the results of MME99 and only marginally inconsistent with 
the BBLP02 models of similar entropy injection.

What about their favored models?  We have tried the same type of test on their S3 
data set ($K_0 \approx 100$ keV cm$^2$).  Fitting all simulated clusters with 
$T_X \gtrsim 2$ keV (mean temperature of 3.8 keV) we find $M_{gas} \propto T_X^{1.86 
\pm 
0.12}$, which is in good agreement with the results of MME99.  When we remove all 
clusters below 3 keV (mean temperature of 4.9 keV), however, the best fit is $M_{gas} 
\propto T_X^{1.77 \pm 0.38}$.  In this case, the best-fit relation is not very 
constraining.  It is even indistinguishable from the self-similar result.  It is 
apparent from their Figure 1, however, that the predicted normalization for this 
model (and all other low entropy models) does not match the observations of MME99.  
This is noted by the authors themselves.  They claim the difference in the zero point 
may be resolved by reducing the baryon fraction by $\sim 20\%$.  As we noted earlier, 
however, a similar normalization offset is also seen in the total cluster mass - 
temperature ($M - T_X$) relation and this cannot be resolved by reducing the baryon 
fraction.  This suggests that the problem lies with the temperature, rather than the 
gas mass.  Alternatively, Bialek et al. also suggest that rescaling their simulations 
for $H_o = 70$ km s$^{-1}$ Mpc$^{-1}$ (instead of 80 km s$^{-1}$ Mpc$^{-1}$) would 
bring consistency between the normalization of this model and the observations.  This 
would be true only if the baryon fraction was held fixed at 0.1 and not rescaled for 
the new cosmology.  Given that they assume $\Omega_m = 0.3$, this would imply 
$\Omega_b = 0.015 h^{-2}$ which is roughly 30\% lower than observed in quasar 
absorption spectra (Burles \& Tytler 1998).  Thus, while the normalization 
offset between their theoretical model and the observations of MME99 is directly 
reduced by decreasing the value of $h$, it is indirectly increased by roughly the 
same proportion through the increased value of $\Omega_b/\Omega_m$.

In summary, as with the Loewenstein (2000) models, we find that the difference 
in the results and conclusions of Bialek et al. (2001) and the present work can be 
explained on the basis that different temperature ranges were examined.  In 
particular, we have shown that the fraction of cool clusters in Bialek et al.'s 
simulated data set is much larger than that found in the MME99 sample and this has 
likely led to an underestimation of the entropy floor in these clusters.  In order to 
safely and accurately compare the preheated models of BBLP02 with observations we 
have paid special attention to only those hot clusters with $T_X \gtrsim 3$ keV.  As 
such, we believe our comparison is more appropriate.

\section{Discussion \& Conclusions}

Motivated by a number of observational studies that have suggested that the $M_{gas} - 
T_X$ relation of clusters of galaxies is inconsistent with the self-similar result of 
numerical simulations and by the launch of the {\it Chandra} and {\it XMM-Newton} 
satellites, which will greatly improve the quality of the observed $M_{gas}-T_X$ 
relation, we have implemented the analytic model of BBLP02 to study the impact of 
preheating on $M_{gas} - T_X$ relation.  The predictions of the model have previously 
been shown to be in very good agreement with observations (e.g., $L_X - T_X$ relation 
and $L_X - \sigma$ relation).

In agreement with the previous theoretical studies of Loewenstein (2000) and 
Bialek et al. (2001), our analysis indicates that injecting the intracluster 
medium with entropy leads to a steeper relationship than predicted by the 
self-similar result of numerical simulations of clusters that evolve through the 
effects of gravity alone.  Loewenstein (2000) and Bialek et al. (2001) have found 
that models that produce an entropy floor of $K_0 \sim 100$ keV cm$^2$, which is 
consistent with measurements of galaxy groups,  are capable of reproducing the 
$M_{gas} - T_X$ relation of hot clusters.  This is inconsistent with our analysis, 
which indicates that a ``high'' level of entropy injection ($K_0 \gtrsim 300$ 
keV cm$^2$) is required to match the observational data of hot clusters of White et 
al. (1997), Peres et al. (1998), and MME99.   It is also inconsistent with BBLP02's 
best-fit value of $K_0 \approx 330$ keV cm$^2$ found via an investigation of the $L_X 
- T_X$ relation of both groups and hot clusters.  They note that the strongest 
constraints for a high entropy floor comes from hot clusters.  Moreover, a high 
value of $K_0$, one that is inconsistent with the  predictions of the best-fit 
models of Loewenstein (2000) and Bialek et al. (2001), has also been reported by 
Tozzi \& Norman (2001).  Finally, da Silva et al. (2001) used numerical simulations 
with a ``low'' value of $K_0 \sim 80$ keV cm$^2$ (which is similar  to 
the predictions of the best-fit models of Loewenstein 2000 and Bialek et al. 2001) 
and found that they {\it could not} reproduce the observed X-ray scaling relations.  
Our result, on the other hand, is consistent with the results of BBLP02, Tozzi \& 
Norman (2001), and da Silva et al. (2001).  As discussed in \S 5, we believe the 
difference between the studies of Loewenstein (2000) and Bialek et al. (2001) and 
present work can be explained by considering the difference in temperature ranges 
studied.  In particular, we have focused only on hot clusters in an attempt to match 
the majority of the observational data as closely as possible.  The results and 
conclusions of the other two studies, however, are strongly influenced by their low 
temperature model data.

We have proposed that the $M_{gas} - T_X$ relation can be used as a probe of the gas 
density profiles of clusters if it is evaluated at different fixed radii.  This is a 
new test.  The preheated models of BBLP02 predict a mild break in the 
scaling relations when small fixed radii (such as $r = 0.250 h^{-1}$ Mpc) are used.  
The scatter in the current observational data is consistent with the predictions of 
the BBLP02 models with $K_0 \gtrsim 300$ keV cm$^2$; however, the exact shape of the 
gas density profiles is not tightly constrained.  We anticipate that large samples of 
clusters observed by {\it Chandra} and {\it XMM-Newton} will place much stronger 
constraints on the gas density profiles of clusters and allow for further testing of 
the preheating scenario.

Finally, the high level of energy injection inferred from our analysis has important 
implications for the possible sources of this excess entropy.  Valageas \& Silk (1999), 
Balogh et al. (1999), and Wu et al. (2000) have all shown that galactic winds driven 
by supernovae can only heat the intracluster/intergalactic medium at the level of 
$\lesssim 0.3 - 0.4$ keV per particle.  This is lower than the $1 - 2$ keV per 
particle result found here.  Thus, if the BBLP02 preheated models provide an accurate 
description of the ICM, supernovae winds alone cannot be responsible for the excess 
entropy.  It has also been speculated that quasar outflows may be responsible (e.g., 
Valageas \& Silk 1999; Nath \& Roychowdhury 2002).  This remains an open possibility.

The role of radiative cooling also remains an open issue.  Recently, it has been 
suggested that {\it both radiative cooling and preheating together} could be actively 
involved in shaping the X-ray scaling relations (e.g., Voit \& Bryan 2001; Voit et al. 
2002).  Radiative cooling (and subsequent star formation) would serve to remove the 
lowest entropy gas, which in turn would help to compress the highest entropy gas, 
thus increasing the emission-weighted gas temperature and steepening the $M_{gas} - 
T_X$ relation (cf. the discussion of entropy in the Cool+SF simulation of Lewis et 
al. 2000).  In this way, the combination of cooling and preheating may reduce the 
best-fit entropy level, perhaps even to a level that can be provided by supernovae 
winds (Voit et al. 2002).  Further study is required to determine the relative roles 
that both preheating and cooling have on cluster evolution.

\noindent {\bf Acknowledgements}

We would like to thank Mike Loewenstein and the anonymous referee for many useful 
comments and suggestions.
I. G. M. is supported by a postgraduate fellowship from the Natural Sciences and 
Engineering Research Council of Canada (NSERC) and by the Petrie Fellowship at the 
University of Victoria.  He also acknowledges additional assistance in the form of a 
John Criswick Travel Bursary.  A. B. is supported by an NSERC operating grant and M. 
L. B. is supported by a PPARC rolling grant for extragalactic astronomy and cosmology 
at the University of Durham.

\newpage

\begin{deluxetable}{lccc}
\tablecaption{Results of linear regression fits to $M_{gas}(r_{500}) - T_X$ data
\label{tab1}}
\tablewidth{35pc}
\tablehead{
\colhead{Model}                      &
\colhead{Entropy Floor (keV cm$^2$)} &
\colhead{m$^a$}        &
\colhead{b$^a$}
}
\tablenotetext{~}{Note.--- We have fit models of the form $\log{M_{gas}} = m 
\log{T_X} + b$ over the range 3 keV $\lesssim T_X \lesssim 10$ keV.}
\tablenotetext{a}{Uncertainties correspond to the 90\% confidence level.}
\startdata
Isothermal model & 0 & 1.50 & 12.72 \\
Preheated  models & 100 & 1.65 & 12.54\\
 & 200 & 1.66 & 12.48\\
 & 300 & 1.67 & 12.45\\
 & 427 & 1.68 & 12.42\\
MME99 data & ? & $1.93 \pm 0.16$ & $12.21 \pm 0.13$\\
MME99 data minus 2 clusters & ? & $1.82 \pm 0.14$ & $12.31 \pm 0.12$\\
\enddata
\end{deluxetable}

\begin{deluxetable}{lccc}
\tablecaption{Results of linear regression fits to $M_{gas}$($r = 0.25 h^{-1}$ Mpc) - 
$T_X$ data
\label{tab2}}
\tablewidth{35pc}
\tablehead{
\colhead{Model} &
\colhead{Entropy Floor (keV cm$^2$)} &
\colhead{m$^a$}        &
\colhead{b$^a$}
}
\tablenotetext{~}{Note.--- We have fit models of the form $\log{M_{gas}} = m \log{T_X} 
+ b$ 
over the range 3 keV $\lesssim T_X \lesssim 10$ keV.}
\tablenotetext{a}{Uncertainties correspond to the 90\% confidence level.}
\startdata
Isothermal model & 0 & 0.84 & $12.61$ \\
Preheated  models & 100 & $0.91$ & $12.43$\\
 & 200 & $0.95$ & $12.34$\\
 & 300 & $1.06$ & $12.21$\\
 & 427 & $1.19$ & $12.03$\\
Peres et al. and White et al. data & ? & $1.11 \pm 0.22$ & $12.06 \pm 0.17$\\
\enddata
\end{deluxetable}

\begin{deluxetable}{lccc}
\tablecaption{Results of linear regression fits to $M_{gas}$($r = 0.50 h^{-1}$ Mpc) -
$T_X$ data
\label{tab3}}
\tablewidth{35pc}
\tablehead{
\colhead{Model}                      &
\colhead{Entropy Floor (keV cm$^2$)} &
\colhead{m$^a$}        &
\colhead{b$^a$}
}
\tablenotetext{~}{Note.--- We have fit models of the form $\log{M_{gas}} = m \log{T_X} 
+ b$ over the range 3 keV $\lesssim T_X \lesssim 10$ keV.}
\tablenotetext{a}{Uncertainties correspond to the 90\% confidence level.}
\startdata
Isothermal model & 0 & 0.97 & 12.87 \\
Preheated  models & 100 & 0.98 & 12.78\\
 & 200 & 0.98 & 12.75 \\
 & 300 & 0.99 & 12.72 \\
 & 427 & 1.01 & 12.68 \\
White et al. data & ? & $1.12 \pm 0.25$ & $12.53 \pm 0.20$\\
\enddata
\end{deluxetable}

\end{document}